\documentclass[conference]{IEEEtran}
\IEEEoverridecommandlockouts
\usepackage{cite}
\usepackage{amsmath,amssymb,amsfonts}
\usepackage{algorithmic}
\usepackage{graphicx}
\usepackage{textcomp}
\usepackage{xcolor}
\usepackage{caption, tabularx}
\usepackage{subcaption}
\usepackage{float}
\usepackage{url}

\def\BibTeX{{\rm B\kern-.05em{\sc i\kern-.025em b}\kern-.08em
    T\kern-.1667em\lower.7ex\hbox{E}\kern-.125emX}}

\usepackage{standalone}
\usepackage{tikz}

\newtheorem{lem}{Lemma}

\begin{document}
\bstctlcite{EE:BSTcontrol}
\nocite{EE:BSTcontrol}
\title{
	 End-to-End NOMA with Perfect and Quantized CSI  Over Rayleigh Fading Channels   \\
}

\author{%
	\IEEEauthorblockN{Selma Benouadah$^{\dagger}$, Mojtaba Vaezi$^{\dagger}$,  Ruizhan Shen$^{\ddagger}$, and Hamid Jafarkhani$^{\ddagger}$  }
	\IEEEauthorblockA{
		$^{\dagger}$Department of ECE,
		Villanova University, Villanova, PA 19085, USA\\
				$^{\ddagger}$Electrical Engineering and Computer Science, University of California, Irvine, CA 92697, USA  \\
		Emails:   \{sbenouad, mvaezi\}@villanova.edu, \{ruizhans, hamidj\}@uci.edu
	}
	\thanks{This work was supported by the U.S. National Science Foundation  under Grants ECCS-2301778 and CCF-2328075.
}
}

\maketitle

\begin{abstract}
An end-to-end autoencoder (AE) framework is developed for downlink non-orthogonal multiple access (NOMA) over Rayleigh fading channels, which learns interference-aware and channel-adaptive super-constellations. While  existing works either assume additive white Gaussian noise channels or treat fading channels without a fully end-to-end learning approach, our framework directly embeds the wireless channel into both training and inference. To account for practical channel state information (CSI), we further incorporate limited feedback via both uniform and Lloyd–Max quantization of channel gains and analyze their impact on AE training and bit error rate (BER) performance. Simulation results show that, with perfect CSI,  the proposed AE outperforms the existing analytical NOMA schemes. In addition, Lloyd–Max quantization achieves superior BER performance compared to uniform quantization. These results demonstrate that end-to-end AEs trained directly over Rayleigh fading can effectively learn robust, interference-aware signaling strategies, paving the way for NOMA deployment in fading environments with realistic CSI constraints.
\end{abstract}

\section{Introduction}

Non-orthogonal multiple access (NOMA) has attracted attention as a next-generation multiple access technique capable of supporting massive connectivity~\cite{jafarkhani2024modulation}. 
In NOMA, multiple users are simultaneously served in the same time–frequency resource by superposing their symbols at the transmitter using different power allocation factors, while successive interference cancellation (SIC) is employed at the receivers. Despite its theoretical benefits, the practical realization of NOMA with finite-alphabet constellations like  quadrature amplitude modulation
 (QAM) is nontrivial, especially under fading conditions where inter-user interference and channel randomness lead to overlapping super-constellations and severe bit error rate (BER) degradation ~\cite{6204010,9552864}.

The majority of studies on NOMA focus on additive white Gaussian noise (AWGN) channels. 
However, the stochastic nature of the Rayleigh fading channel introduces additional challenges, including variable power ordering. 
Under Rayleigh fading, most existing works assume Gaussian inputs and analyze achievable rates and outage probabilities~\cite{8352126,9497653,9598834,9802725,9348935}, 
while only a few consider BER or symbol error rate (SER) as a performance metric~\cite{8449119,9075282,9403384} and propose analytical approaches to improve it. 
Furthermore, with a few exceptions~\cite{XLHJ17,XZMGHJwcl20}, most studies on downlink NOMA assume perfect channel state information (CSI) at the transmitter.

Recently, deep learning techniques have been employed to enhance the performance of NOMA systems~\cite{ye2021deep,van2020deep,8580937,10304541,vaezi2025AENOMA}.
In particular, several studies have demonstrated the potential of autoencoder (AE)–based end-to-end NOMA transmission over AWGN channels~\cite{vaezi2025AENOMA,8580937,10304541}.
AE-based systems are capable of jointly learning the transmitter and receiver mappings, achieving remarkable BER performance~\cite{8580937,10304541,vaezi2025AENOMA}.
In particular, with a properly designed loss function, they can learn interference-aware super-constellations instead of relying on SIC~\cite{vaezi2025AENOMA}.

The AE-based NOMA schemes discussed above have been investigated only under AWGN channels, and their extension to Rayleigh fading remains unexplored. The stochastic nature of fading requires the encoder to adapt to random CSI across users, unlike in AWGN channels where a single fixed mapping suffices. Rayleigh fading further demands optimization over the fading distribution, substantially increasing training complexity. Moreover, existing AE-NOMA designs generally assume perfect CSI at the transmitter.  

This work develops an AE-NOMA over Rayleigh fading channels, incorporating both perfect and quantized CSI feedback, with BER/SER as the primary performance metrics. To the best of our knowledge, this is the first study to address AE-NOMA under Rayleigh fading with quantized CSI. The main contributions are summarized as follows:

\begin{itemize}
\item We extend AE-NOMA from AWGN to Rayleigh fading, where the encoder adaptively shapes the super-constellation to mitigate inter-user interference, ensuring good BER performance across all channel gains.
\item To make the design more practical, we incorporate quantized CSI feedback using both uniform and Lloyd–Max schemes, resulting in the first AE-NOMA framework with quantized CSI.
\item We propose a novel training and loss function to mitigate the BER floor observed at high SNR.
\end{itemize}
Our extensive numerical results compare the average BER with the theoretical BER derived for quadrature phase-shift keying (QPSK) transmission for both users, as well as with the results in [12] and [14], which employ adaptive power allocation and constellation rotation. The proposed AE-NOMA scheme noticeably outperforms both benchmarks.
 
The remainder of this paper is organized as follows. Section~II presents the system model. Section~III introduces the proposed AE architecture and training approach. Section~IV provides numerical results and Section~V concludes the paper.

\section{System Model}
We consider the general problem of downlink power domain NOMA consisting of a single base station (BS) and two users. The BS maps binary input bits $\mathbf{s}_1$ and $\mathbf{s}_2$ to complex-valued symbols $x_1$ and $x_2$, and
superimposes them onto a signal  
\begin{equation}
x = \sqrt{\alpha P}x_1 + \sqrt{(1- \alpha)P}x_2,
\label{Eq:noma1}
\end{equation}
 and transmits it over $n$ channel uses,  where  $P$ is the maximum transmit power and $\sqrt{\alpha P}$ and $\sqrt{(1- \alpha)P}$ are the powers allocated to Users 1 and 2, respectively. 

We consider a set of independent \textit{Rayleigh fading} channels between the BS and each of the two users. The signal received at user equipment $k$ (UE$k$), $k \in \{1,2\}$, is represented by
\begin{equation}
y_k = h_kx+n_k,
\label{Eq:channel}
\end{equation}
\noindent where $h_k =h_{r_k}+ ih_{i_k}$,  with  $h_{r_k},  h_{i_k} \sim \mathcal{N}(0,\sigma_{h_k}^{2})$, is the channel and $n_k \sim \mathcal{CN}(0,\sigma_k^{2})$ is Gaussian noise. Therefore, the channel gain $|h_k|=\sqrt{h_{r_k}^2+h_{i_k}^2}$ has a Rayleigh distribution.  At the receiver side, an estimated information sequence $\hat{x}_k$ is obtained from decoding the received signal $y_k$,  $k \in \{1,2\}$. 

\section{Autoencoder for NOMA Rayleigh Channel}
\label{sec:ae_noma}
We propose an autoencoder that jointly optimizes the transmitter and receivers through end-to-end learning, considering both perfect and quantized CSI. 
The structure of the network is illustrated in Fig.~\ref{fig:netModel}. 
The objective is to transmit binary input messages $\mathbf{s}_1$ and $\mathbf{s}_2$ of lengths $\ell_1$ and $\ell_2$, respectively.

\subsection{Network Structure and Training}
 The AE comprises an encoder representing the transmitter and two decoders corresponding to UE1 and UE2. The encoder maps binary input $\mathbf{s}_1$ and $\mathbf{s}_2$ to a complex-valued superimposed symbol that satisfies the average power constraint $P$. The detector $k$ maps its received noisy observation back to estimated bits $\hat{\mathbf{s}}_k$.
Unlike a conventional communication system with fixed constellations, the AE learns a nonlinear mapping based on the instantaneous $h_1$ and $h_2$, or their quantized values $\hat h_1$ and $\hat h_2$. In contrast, each decoder utilizes only its corresponding channel gain as an input. By including channels as an input to the encoder, the transmitter can adaptively shape the superimposed constellation in response to current channel conditions.

In the AE-NOMA design for the AWGN channel~\cite{vaezi2025AENOMA,8580937,10304541}, the channel coefficients $h_1$ and $h_2$ are constant, resulting in $2^{\ell_1 + \ell_2}$ possible super-constellations corresponding to $2^{\ell_1 + \ell_2}$ input vectors.
In contrast, under Rayleigh fading, the channel gains vary with each transmission, and the input to the AE effectively represents super-constellation  over all possible combinations of channel gains for UE1 and UE2.
This significantly increases the complexity of mapping the inputs to the super-constellation.  

\begin{figure*}[htbp]
    \centering
    \includegraphics[width=0.9\textwidth]{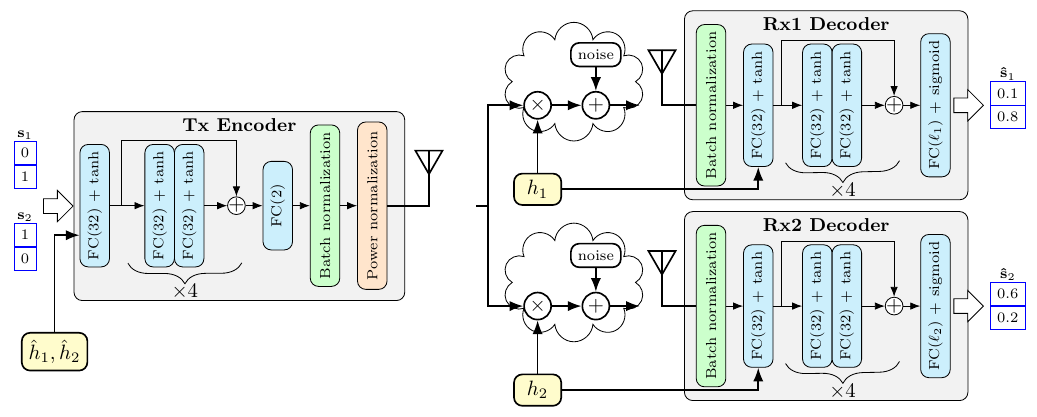}
    \caption{The architecture of the implemented AE-NOMA.   In this example, the output layer sizes  of the decoders are $\ell_1=\ell_2=2$. In the full CSI case, $\hat{h}_k = h_k$, whereas in the quantized CSI case, $\hat{h}_k = f_Q(h_k)$ for $k \in \{1,2\}$.
}
    \label{fig:netModel}
\end{figure*}

{
The network architecture is shown in Fig.~\ref{fig:netModel}. 
The encoder and decoders each consist of an input layer, eight hidden layers, and an output layer. 
All layers are fully connected (FC) but have different numbers of neurons, as indicated in the figure. 
Compared to the AWGN case~\cite{vaezi2025AENOMA}, 
a deeper architecture with eight hidden layers is adopted for the Rayleigh case to better capture the mapping induced by varying channel conditions. 
The input and hidden layers each contain 32 neurons and are followed by a $tanh$ activation function. 
The encoder’s output layer contains two neurons for the Tx, representing the \textit{in-phase} (I) and \textit{quadrature-phase} (Q) components. 
The decoders’ output layers have $\ell_1$ and $\ell_2$ neurons, respectively, corresponding to the number of bits for each user. 
Each decoder outputs the bit probabilities through a \textit{sigmoid} activation function. 
Residual connections in the Tx, Rx1, and Rx2 networks are included to preserve gradients and enhance learning performance.
}

The end-to-end AE is trained to minimize a weighted sum 
\begin{equation}
\mathcal{L}(\gamma) = w \max(\mathcal{L}_1, \mathcal{L}_2) + \min(\mathcal{L}_1, \mathcal{L}_2),
\label{Eq:loss_f}
\end{equation}
where $\mathcal{L}_1$ and $\mathcal{L}_2$ denote the \textit{binary cross-entropy} losses for UE1 and UE2, respectively, 
$w$ is a weighting factor, and $\gamma$ represents the SNR value used during training. 
By setting $w>1$, the loss function emphasizes the user with the higher loss, thereby enforcing user fairness.
The AE is implemented in PyTorch using the Adam optimizer with learning rate decaying with an exponential schedule for better convergence and performance.

\subsection{Quantized CSI versus Perfect CSI}
In practical systems, the BS cannot always access perfect CSI due to feedback bandwidth limitations. To emulate realistic constraints, quantization is applied to the instantaneous channel amplitudes before feeding them to the AE encoder. 
For the results presented in the next section, we assume full CSI at the receivers and, at the transmitter, either full CSI (when no quantization is applied) or quantized CSI.

We separately quantize the real and imaginary parts of the channel gains. 
We consider a scalar quantizer with $M$ output levels 
and decision boundaries
$b_i$ ($b_0=-\infty,\,b_M=+\infty$) that maps $h$ to $\hat h$ based on
\[
\hat h = f_Q(h) \quad \text{if } b_{i-1}<h\le b_i,
\]
and its distortion is measured by the mean squared quantization error (MSQE)
$\mathbb{E}[(h - f_Q(h))^2]$. 
We employ both uniform and Lloyd--Max (LM) quantizers \cite{gray1998quantization}. 
Uniform quantization fixes the decision boundaries and reconstruction levels to be equally spaced over a given support. 
It does not adapt to the source probability density function (pdf) and is optimal only for a uniformly distributed source. 
In contrast, Lloyd--Max quantization minimizes the MSQE by iteratively updating the centroids and boundaries as functions of the source pdf~\cite{gray1998quantization}. 
Unlike uniform quantization, Lloyd--Max allocates finer resolution where the pdf  is more concentrated, thereby achieving significantly lower distortion.


\section{Numerical Results}
This section presents the numerical results obtained for the proposed AE architecture under Rayleigh fading channels. All tests were conducted (unless otherwise mentioned) with $40{,}000$ training samples, $400{,}000$ testing samples, and $10{,}000$ training epochs. To ensure statistical robustness \cite{dutta2022seed}, since each seed may result in slightly different results, all tests were performed for several seeds, and the reported BER curves were averaged over multiple random seeds. The training SNR is 10\,dB, the weight loss is $w=10$, and we conduct training and testing with varying channel gains drawn from the Rayleigh distribution with the fading parameter $\sigma_{h_1}=1$ for the weak user (UE1) and $\sigma_{h_2}=2$ for the strong user (UE2). The number of bits of each user is $\ell_1=\ell_2=2$ bits. The training with exponential decay schedule starts from a learning rate of $0.01$ and decays each $100$ epochs by a drop of $0.95$. The results in the figures present $(\rm BER_1+BER_2)/2$, the average BER of the two users.

\subsection{Super-constellation Adaptation to Channel Gains}
\begin{figure}[h!]
\centering
\begin{subfigure}{0.49\linewidth}
\centering
    \includegraphics[width=0.99\linewidth]{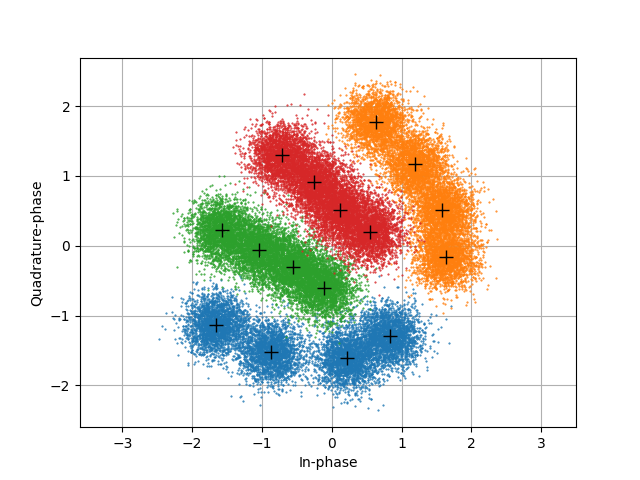} 
    \caption{UE1 ($h_2=2h_1=2$)}
\end{subfigure}
\hfill
\begin{subfigure}{0.49\linewidth}
\centering
    \includegraphics[width=0.99\linewidth]{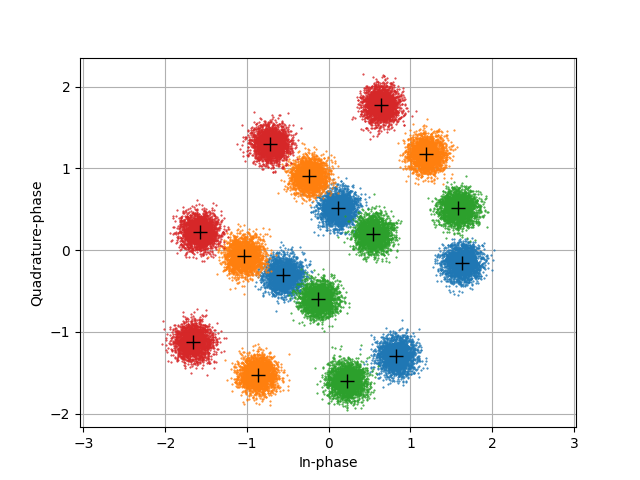} 
    \caption{ UE2 ($h_2=2h_1=2$)}
\end{subfigure}
\hfill
\begin{subfigure}{0.49\linewidth}
\centering
    \includegraphics[width=0.99\linewidth]{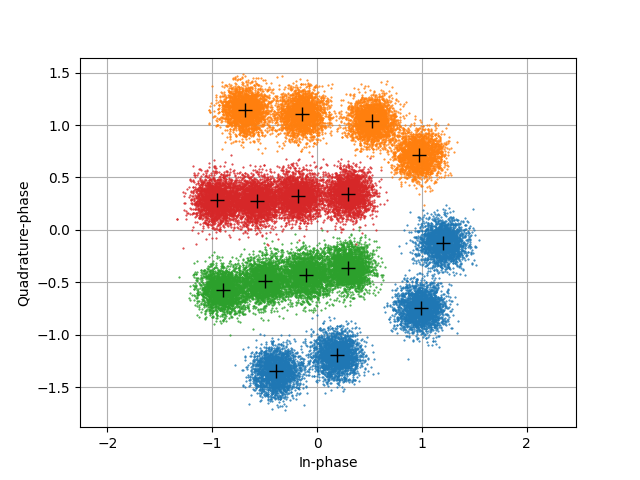} 
    \caption{ UE1 ($h_2=2h_1=2+2j$)}
\end{subfigure}
\hfill
\begin{subfigure}{0.49\linewidth}
\centering
    \includegraphics[width=0.99\linewidth]{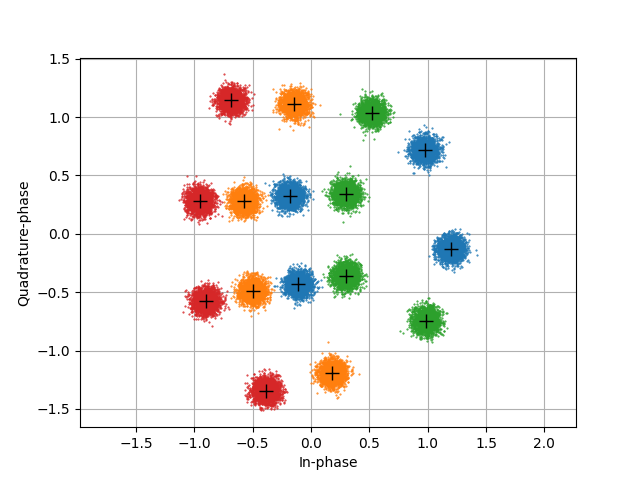} 
    \caption{UE2 ($h_2=2h_1=2+2j$)}
\end{subfigure}
\caption{ Super-constellations with two different channel gains for UE1 and UE2.}\label{fig:res_sc}
\end{figure}

Conventional NOMA designs generate super-constellations with fixed power allocation and preset constellations for UE1 and UE2 (e.g., QPSK). Prior works, such as \cite{8449119}, improved performance by rotating one user’s constellation before superposition and adapting the power allocation and rotation based on instantaneous CSI, showing that both strategies enhance NOMA reliability. Building on this, our AE operates without preset constellations or fixed allocations, allowing it to explore an unlimited set of super-constellation structures. Additionally, power allocation is not fixed across all bit combinations so each combination of bits for UE1 and UE2 can have different allocation factors, with only an average power constraint applied to the overall super-constellation.
Figure~\ref{fig:res_sc} illustrates two instances of designed constellations. Each point represents a received symbol, with colors indicating the same message for a given user. The weaker user (UE1) observes four clusters, while the stronger user (UE2) can resolve up to sixteen clusters. The constellation structure changes with the channel.

\subsection{Comparison with Theoretical and Benchmark BER Bounds}

To compare the performance of the proposed AE-NOMA, we first derive the theoretical SER for a two-user NOMA over a Rayleigh fading channel, where each user employs QPSK modulation. The result is provided in Lemma~\ref{lemma} in Appendix~\ref{sec:ber_noma}.  Figure~\ref{fig:ber_bounds} presents this comparison for a power allocation factor of $\alpha = 0.7$ for UE1. Across the considered SNR range, the AE consistently outperforms the theoretical BER when a QPSK–QPSK constellation is applied. This confirms the effectiveness of the learned modulation and decoding strategy.
\begin{figure}[t]
\centering
\includegraphics[width=0.8\linewidth]{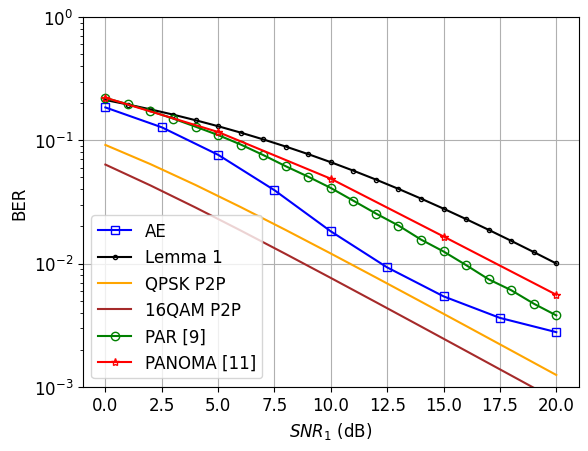}
\caption{Average BER performance of the proposed AE system and traditional approaches under Rayleigh fading.}
\label{fig:ber_bounds}
\end{figure}
For further insight, we include the Rayleigh BER performance of a  \textit{point-to-point}  16-QAM \cite{goldsmith2005wireless} constellation with Rayleigh distribution parameter $\sigma_{h_2}=2$, motivated by the fact that, from the perspective of UE2, an ideal AE-learned super-constellation should resemble a 16-QAM (or a rotated variant), as this structure maximizes the minimum Euclidean distance and can serve as a practical theoretical lower bound for UE2 since it must be able to distinguish all symbols. 
We include the Rayleigh BER of QPSK single–user link as a lower bound for UE1, with Rayleigh distribution parameter $\sigma_{h_1}=1$, since UE1 should experience a clean QPSK channel. 

Our learned AE-based constellation at UE1 closely tracks the lower bound. We also observe that the BER of the AE-designed constellation indeed approaches the 16-QAM benchmark, particularly around the training SNR of 10\,dB, suggesting that the AE learns a near-optimal constellation structure.  At a BER of $10^{-2}$, 16-QAM achieves this target at approximately 9\,dB, whereas the AE reaches it at around 12.5\,dB, better than the theoretical NOMA baseline, which requires nearly 17.5\,dB.

We also benchmark against two adaptive downlink NOMA baselines proposed in \cite{8449119} and \cite{9403384}, referred to as PAR and PANOMA, respectively, in Fig.~\ref{fig:ber_bounds}. In \cite{8449119}, the authors improve the BER of downlink NOMA using an adaptive power allocation scheme and rotation of the stronger user's constellation. The scheme proposed in \cite{9403384} is an adaptive power factor scheme. The authors divide the symbols of the two users' constellations into two parts: symbols with fixed power factor regardless of the channel conditions (identical symbols from the two constellations) with $0.5$ factor for UE1 and UE2, while the power for the remaining symbols can be selected to minimize the average BER. The power factor of the remaining symbols is found through simulation. The authors provide analytical results for BPSK case while simulation based results only for QPSK case. The optimal factor for QPSK case is 0.95.

Compared to both \cite{8449119} and \cite{9403384}, our AE model outperforms the BER performance across the full SNR range. The key difference is that prior works assume QPSK constellations as the base signaling scheme, while our approach learns interference-aware super-constellations without enforcing predefined constellation structures.



\subsection{Adaptability for Different Bit Lengths}

\begin{figure}[t]
\centering
\includegraphics[width=0.8\linewidth]{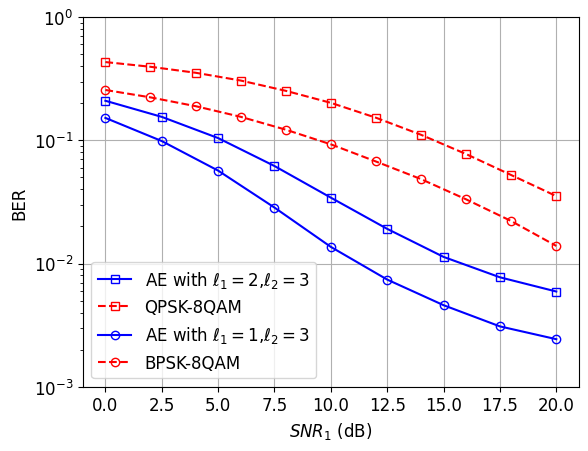}
\caption{Average BER performance of AE compared to traditional constellations for two cases: $\ell_1=2$ bits, $\ell_2=3$ bits and $\ell_1=1$ bit, $\ell_2=3$ bits.} \label{fig:res_2_3_sim}
\end{figure}

We consider scenarios with different numbers of bits per user to demonstrate the adaptability and scalability of the proposed network. Two cases are evaluated: Case 1, where UE1 transmits 2 bits and UE2 transmits 3 bits (QPSK–8QAM); and Case 2, where UE1 transmits 1 bit and UE2 transmits 3 bits (BPSK–8QAM). Since no closed-form expression exists as in Lemma~\ref{lemma}, we simulate the BER performance of NOMA with conventional constellations over Rayleigh fading, as shown in Fig.~\ref{fig:res_2_3_sim}. In these simulations, the power allocation factor is set to $\alpha = 0.9$, which yields the lowest BER and was determined via an exhaustive search.

The AE adapts to changes in the number of bits without any modification to its overall structure, except for adjusting the decoder output sizes to match the bit lengths of each user. The proposed AE-based NOMA significantly outperforms conventional NOMA in both cases. This improvement arises because, as the number of bits increases, the combination of two traditional constellations leads to a higher likelihood of overlapping symbols.

\subsection{Quantized CSI Feedback: Uniform vs.~Lloyd–Max}
\begin{figure}[t] \centering \includegraphics[width=0.8\linewidth]{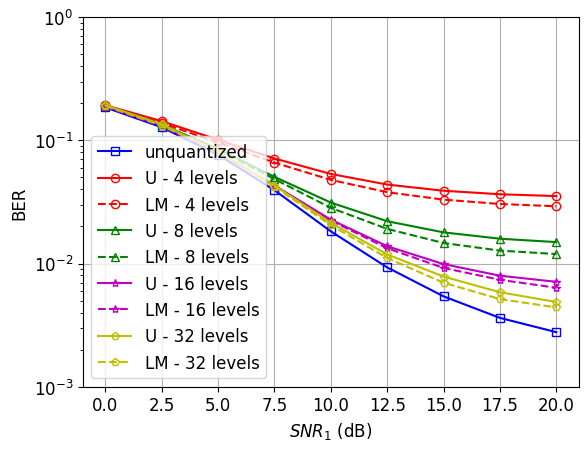}  \caption{Average BER for uniform (U) and Lloyd–Max (LM) quantized CSI feedback under various quantization levels and unquantized CSI. All graphs are for $l_1=l_2=2$ bits.}\label{fig:quant_uniform} \end{figure}

We evaluated two quantization schemes for limited feedback: uniform and Lloyd–Max quantization. Figure~\ref{fig:quant_uniform} illustrates the corresponding BER curves. We first evaluate the impact of feedback resolution under quantization. As expected, increasing the number of quantization levels consistently improves the BER performance for both users. Compared to uniform quantization, Lloyd–Max consistently achieves lower BER across all levels, since it minimizes quantization error. At higher resolutions ($\geq 4$ bits), the performance gap between the two schemes becomes small. This suggests that overall AE performance could be further improved by jointly learning power allocation and quantization centroids, rather than relying on predefined scalar quantizers. 

\subsection{Improving BER Floor at High SNR}
A slight BER floor is observed in the AE-based NOMA scheme at high SNR in Fig.~\ref{fig:ber_bounds}. This effect is not due to noise, but rather a consequence of the fixed training SNR used during optimization, also observed in point-to-point communications \cite{song2022benchmarking}. Higher training SNR push the floor rightward and can remove the floor. However, if the training SNR is set too high, the AE performance becomes worse at low SNRs, leading to degraded performance in the practical operating region. In this sense, training at 10\,dB provides a good balance for NOMA, avoiding early saturation at high SNR while preserving strong performance at low SNRs where reliability is most critical. 

\begin{figure}
    \centering
\includegraphics[width=0.8\linewidth]{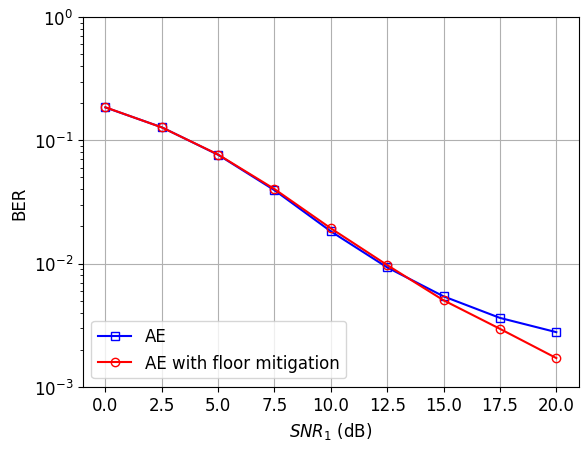}

    \caption{Average BER performance of  AE  before and after flooring training mitigation strategy for unquantized CSI.}
    
    \label{fig:floor_mit}
\end{figure}

To address this flooring issue, we propose a new training method. We train each batch with several representative SNR values and calculate the conditional weighted loss of the users $\mathcal{L}(\gamma)$ in (\ref{Eq:loss_f}) for each. Then, we apply an additional weight to the $\mathcal{L}(\gamma)$, which is based on the normalized SNR used for training in this iteration over the sum of the representative SNRs used for training. The overall loss function designed for each batch, $\mathcal{L}_{F}$, is expressed as
\begin{equation}
     \mathcal{L}_{F} = \sum_{i=1}^{|\mathcal S|} \frac{\gamma_i}{\sum_{j=1}^{|\mathcal S|}\gamma_j} \mathcal{L}(\gamma_i),
\end{equation}
\noindent where $\gamma_i$ is the SNR used during the training from the set of representative SNRs $\mathcal S$ with cardinality $|\mathcal S|$. With this training method, the floor performance is mitigated while no significant effect on low SNR is observed, as illustrated in Fig.~\ref{fig:floor_mit}. In addition, the training complexity is reduced as only five representative SNRs were needed for training, which are $\mathcal S=\{1,5,10,15,20\}$ instead of going through all the values randomly in the range.

Overall, the proposed AE architecture for NOMA exhibits strong generalization under practical Rayleigh fading with power and feedback constraints, outperforming conventional schemes. The system remains robust under quantized CSI feedback when enough bits are used for quantization, with Lloyd–Max quantization providing measurable gains over uniform schemes, reflecting the impact of mean square quantization error on AE performance.

\section{Conclusion}
We have designed an end-to-end AE-based NOMA structure over a time-varying Rayleigh fading channel, with both perfect and quantized CSI. Through joint optimization of the encoder and decoders, the proposed structure learns adaptive super-constellations for each channel gain. This flexibility allows the network to discover nonconventional symbol mappings that outperform traditional designs. Simulation results confirm that the proposed method achieves significantly better BER performance than conventional NOMA approaches. We have also proposed a new training method to mitigate floor around high SNR, improving the BER and approaching the lower bounds. We have further investigated limited feedback using both uniform and Lloyd–Max quantizers. 
However, the quantizers were not jointly optimized with the AE. 
Future work will explore training of the AE with the quantizer to further close the gap to the unquantized BER performance while reducing the number of levels used for quantization.

\appendix

\subsection{BER for QPSK-QPSK NOMA under Rayleigh Fading}
\label{sec:ber_noma}
Following the standard procedure of \cite{goldsmith2005wireless}, the average error rate of any signaling scheme over a fading channel is obtained by averaging the instantaneous AWGN error probability over the distribution of the instantaneous SNR:
\begin{equation}
\label{eq:avg_ser_generic}
\bar{P}_s \;=\; \int_{0}^{\infty} P_s(\gamma)\, p_{\gamma}(\gamma)\, d\gamma,
\end{equation}
where \(P_s(\gamma)\) denotes the symbol error probability in AWGN for instantaneous SNR \(\gamma\) and \(p_{\gamma}(\gamma)\) denotes the distribution of \(\gamma\) induced by the fading process, which in our case is a Rayleigh channel.
The magnitude of the channel coefficients $|h_k| = \sqrt{h_{r_k}^2 + h_{i_k}^2}$ is Rayleigh distributed with parameter \(\sigma_{h_k}\), with $k \in \{1,2\}$.
Consequently, the instantaneous power \(|h_k|^2\) is exponential with mean $2\sigma_{h_k}^2$. If the instantaneous SNR is proportional to the channel power via \(\gamma=a\,|h_k|^2\), where \(a\) is a non-random scaling factor, then by the scaling property of the exponential distribution, $\gamma=a\,|h_k|^2$ is exponential with mean SNR $2a\sigma_{h_k}^2$ \cite{devore2012modern}.

We consider a two-user downlink NOMA system employing QPSK constellations and SIC. Assuming equal energy per bit for each constellation, denoted \(E_b=P/2\) with \(P\) transmit power, the AWGN SER formulas $P_{s,1}$ and $P_{s,2}$ for UE1 and UE2 respectively, are given by \cite{8288077} 
\begin{subequations}
\label{eq:ps_all}
\begin{align}
P_{s,k} &= 1 - \big(1 - P_{s,k}^I\big)^2, \qquad \quad k \in \{1,2\}, \label{eq:ps_composite_a} \\[4pt]
P_{s,1}^I &= \tfrac{1}{2} \!\left[ Q\!\left( \sqrt{\xi_{1,1}} \right) + Q\!\left( \sqrt{\xi_{1,2}} \right) \right], \label{eq:ps_ue1_awgn_b} \\[4pt]
P_{s,2}^I &= Q\!\left( \sqrt{\xi_{2,1}} \right) + \tfrac{1}{2} Q\!\left( \sqrt{\xi_{2,2}} \right), \label{eq:ps_ue2_awgn_c}
\end{align}
\end{subequations}
\noindent
where  each $\xi_{k,j}$ follows an exponential distribution and are given by
\begin{subequations}
\begin{align}
\xi_{1,1} &= \frac{|h_1|^2 E_b}{\sigma_1^2} \left(\sqrt{1-\alpha} + \sqrt{\alpha}\right)^2, \label{eq:gamma_a}\\
\xi_{1,2} &= \frac{|h_1|^2 E_b}{\sigma_1^2} \left(\sqrt{1-\alpha} - \sqrt{\alpha}\right)^2, \label{eq:gamma_b}\\
\xi_{2,1} &= \frac{|h_2|^2 \alpha E_b}{\sigma_2^2}, \label{eq:gamma_c}\\
\xi_{2,2} &= \frac{|h_2|^2 E_b}{\sigma_2^2} \left(\sqrt{1-\alpha} - \sqrt{\alpha}\right)^2, \label{eq:gamma_d}
\end{align}
\end{subequations}
\noindent and $P_{s,k}^I$ is the probability of User k in the in-phase dimension and the symmetry of the super-constellation in the in-phase and quadrature was exploited to get the SER. 

 To average expressions of the form \(Q\big(\sqrt{a\xi}\big)\) over an exponential SNR we use Craig's integral representation for the \(\mathrm{Q}\)-function $Q(x)= \frac{1}{\pi}\int_{0}^{\pi/2}\exp (-x^{2}/({2\sin^{2}\theta}))\,d\theta$, \cite{goldsmith2005wireless}  then the negative moment generating function or Laplace transform of exponential distribution ($M_{\xi}(s) = \mathbb{E}[e^{-s\xi}] = \frac{1}{1 + s \bar{\xi}}$ for \(\xi\) exponentially distributed with mean \(\bar{\xi}\)). Finally, calculating   $\bar{P}_{s,k}^I =  \int_0^\infty P_s(\xi_k) p_{\xi_k}(\xi_k) d\xi_k$ we get the following Lemma. 
\begin{lem} \label{lemma}
The average symbol error probability of a two-user NOMA system over a Rayleigh fading channel, where each user employs a QPSK constellation, is given by
\begin{align}
\bar{P}_{s,k} &= 1 - \big(1 - \bar{P}_{s,k}^I\big)^2,  \qquad \quad k \in \{1,2\}, \label{eq:ps_avg_c}
\end{align}
in which
    \begin{subequations}
\label{eq:ps_avg_all}
\begin{align}
\bar{P}_{s,1}^I &= \frac{1}{4}\!\left(
2 - \sqrt{\frac{\bar{\xi}_{1,1}}{2+\bar{\xi}_{1,1}}}
- \sqrt{\frac{\bar{\xi}_{1,2}}{2+\bar{\xi}_{1,2}}}
\right), \label{eq:ps_avg_a} \\[4pt]
\bar{P}_{s,2}^I &= \frac{1}{2}\!\left(1 - \sqrt{\frac{\bar{\xi}_{2,1}}{2+\bar{\xi}_{2,1}}}\right)
+ \frac{1}{4}\!\left(1 - \sqrt{\frac{\bar{\xi}_{2,2}}{2+\bar{\xi}_{2,2}}}\right), \label{eq:ps_avg_b} 
\end{align}
\end{subequations}
and average \(\bar{\xi}_{k,j}\) are obtained by replacing \(|h_k|^2\) with its expectation \(2\sigma_{h_k}^2\) in the corresponding instantaneous expressions in \eqref{eq:gamma_a}--\eqref{eq:gamma_d}. 
\end{lem}
The bit error probability under Gray coding is \(P_{b,k} = \tfrac{1}{2}P_{s,k}\).

 \typeout{}

\bibliographystyle{IEEEtran} 
\bibliography{bib}

\end{document}